\documentclass{optica-article}
\usepackage{opticajournal}

\articletype{Research Article}

\usepackage{lineno}
\usepackage{float}

\begin{document}

\title{Fabrication of Customized Low-Loss Optical Resonators by Combination of FIB-Milling and CO$_2$ Laser Smoothing}

\author{Patrick Maier,\authormark{1} Simon Rupp\authormark{2}, Niklas Lettner,\authormark{1} Johannes Hecker Denschlag\authormark{2} and Alexander Kubanek\authormark{1*}}

\address{\authormark{1}Institute for Quantum Optics, Ulm University, 89081 Ulm, Germany}
\address{\authormark{2}Institut für Quantenmaterie and Center for Integrated Quantum Science and Technology (IQST), Ulm University, 89069 Ulm, Germany}
\email{\authormark{*}alexander.kubanek@uni-ulm.de}

\begin{abstract*} 
Fabry-Perot cavities are essential tools for applications like precision metrology, optomechanics and quantum technologies. A major challenge is the creation of microscopic spherical mirror structures which allow the precise matching with the wavefront of a Gaussian beam, while providing high surface quality. 
We present a novel fabrication technique, enabling the creation of customized microscopic cavity mirror structures over a wide range of geometrical parameters, by combining focused ion beam milling (FIB) and CO$_2$ laser smoothing. While FIB milling allows us to imprint features on the mirror substrate with a resolution on the nanometer scale, the application of defocused CO$_2$ laser pulses
consistently reduces remaining surface deformations down to a roughness of $\sigma_\text{RMS}=0.2\,$nm. The average deviation of the profile from a spherical shape is kept below a few nanometres.  
This technique enables the customized and repeatable fabrication of low loss optics on a wide range of optical substrates, including optical fibres. Thus, Fabry-Perot cavities can be fabricated with pre-defined modal volume, high finesse and tailored ellipticity. Since the structural shape of the mirror is created by FIB milling, its pixel-by-pixel nature of the writing process can be used to create arbitrary structures with a resolution given by the FIB milling. At the same time, the surface quality is given by the CO$_2$ laser smoothing processes, without degrading the FIB milled shape.  

\end{abstract*}

\section{Introduction}
Microscopic optical cavities have emerged as pivotal components in numerous scientific and technological applications, ranging from precision metrology \cite{D2LC00180B,Leroux:10,Mader2015} over optomechanics \cite{Hunger:21,Tenbrake2024} to quantum technologies. In particular, experiments in cavity quantum electrodynamics investigate the interaction between single photons and individual quantum emitters like color -centres in solid state materials \cite{Haubler:19,Jensen_2020,Haubler:21,Bayer2023,Feuchtmayr:23,Pallmann2024,Casabone2021,Husel2024,Zifkin:2024,Janitz:20, Herrmann_2024}, quantum dots \cite{Englund2007,Sanchez:2013,Najer2019} and trapped atoms \cite{Haas:2014,Rempe:15,Hartung:2024} or ions \cite{Buifmmode:1997,Steiner:13,Ballance:2017}. The cavity enables various quantum mechanical techniques like cavity-assisted neutral atom traps \cite{Birnbaum:2005,Puppe:2007}, cavity cooling \cite{Maunz2004, Muecke2010, Kubanek:2011} or the controlled formation of molecules \cite{Kampschulte:2018}. Fabrication of mirror structures on end facets of optical fibres has led to a miniaturized footprint \cite{Gulati2017,Pfeifer2022, Brand:2013}. \\
Many experiments require a strong coupling between light and matter, which scales with the ratio between quality factor $Q$ and modal volume of the optical mode $V_m$ of the cavity. Therefore precise and independent control over the cavity geometry and surface topography is required. Up to date methods for manufacturing microscopic cavity optics utilize mechanisms like laser based ablation \cite{Muller:10, Hunger:2012, Brand:2013, Takahashi:14, Ott:16, Hunger2010}, reactive ion etching (combined with laser based ablation) \cite{Najer17}, hydrofluoric acid etching \cite{Peng:19} or dry etching followed by oxidation smoothing \cite{Wachter2019} yielding a finesse $\mathcal{F}$ of up to half a million. Focused Ion Beam (FIB) milling enables modal volumes as low as a few femtoliters \cite{Dolan:10, Kelkar_2015}. \\
Current fabrication techniques however often lack either in surface quality or free choice of structure shape parameters. FIB milling for example is a powerful tool for the arbitrary creation of shapes with features on the nanometer scale, but is limited by surface quality \cite{Trichet15}. CO$_2$ laser based ablation techniques on the other hand show excellent surface quality, but tend to produce Gaussian shaped depressions with a spherical profile mostly in the center of the structure \cite{Hunger2010}, eventually leading to complex mode mixing and increased susceptibility towards misalignment \cite{Hughes:23}. Another disadvantage is that the geometrical parameters of the structure, like the radius of curvature (ROC), the structure depth $h$ and the profile diameter $d$ can not be chosen independently from each other \cite{Hunger:2012}. For example structures with small ROC require a higher structure depth $h$ compared to structures with larger ROC, making a minimisation of the modal volume $V_m$ more difficult \cite{Hunger:2012}. \\
Combination of FIB milling and CO$_2$ laser smoothing \cite{Simsek2017, Walker2018}, allows us to exploit the advantages of each technique. First, spherical concave structures are FIB milled into silica with precision on the nm scale. The free choice of the parameters ROC and structure depth $h$ (which determine the profile diameter $d$) allows minimization and tailoring of the modal volume $V_m$. At the same time high coupling efficiencies to Gaussian modes are feasible. Remaining surface deformations, which introduce scattering losses $\mathcal{S}$, are reduced with CO$_2$ laser smoothing to a level, where the maximal finesse $\mathcal{F}$ is limited mostly by the mirror coating. The theoretical finesse given by the mirror coating, neglecting all other loss channels, is considered the coating limit $\mathcal{F}_{Coating}$. \\
The fabrication procedure is repeated for multiple sets of concave micro cavity arrays on various glass substrates as well as optical fibres. Atomic Force Microscopy (AFM) and optical interferometric methods are employed to evaluate the surface quality and shape of the structures. We compare structural shape and surface topography after each of the two main fabrication steps and estimate optical losses. Finally our method is benchmarked by employing a highly reflective mirror coating and comparing the finesse $\mathcal{F}$ of CO$_2$ laser smoothed and un-smoothed FIB milled structures. \\
Our fabrication process yields concave depressions with an average deviation from a spherical shape of a few nm, while offering the possibility to choose ROC and structure depth $h$ independently from each other over a large parameter space. The surface roughness is consistently reduced to $\sigma_\text{RMS}=0.2\,$nm. The optical properties are examined in a cavity configuration with a finesse of up to $\mathcal{F}\approx 115\,000$.

\section{Customized Cavity Optics}
\label{gauss}

\begin{figure}[h]
\centering
\includegraphics[scale=1]{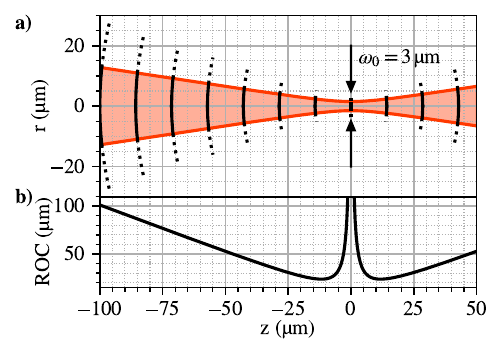}
\caption{\textbf{a)} Schematic representation of a Gaussian beam with a wavelength $\lambda=650\,$nm and a beam waist diameter $\omega_0 = 3\,$µm propagating in free space along an optical axis in $z$-direction. The beam waist $\omega(z)$ (radial distance r from the optical axis, where $I(r)=I_0e^{-1}$) is plotted in red. The radius of curvature (ROC) of the wavefront for different positions along the optical axis is marked in black. \textbf{Inset:} Cavity optics are matched to the Gaussian beam, forming a resonator consisting of two concave mirrros with different radii of curvature (ROC$_1$ and ROC$_2$) at positions $z_1$ and $z_2$. \textbf{b)} Radius of curvature $R(z)$ of the wavefront along the optical axis.}
\label{MatchedGaussian}
\end{figure}

The most commonly used optical mode is the Gaussian TEM$_{00}$-mode. Its free-space propagation is depicted in fig. \ref{MatchedGaussian} a). The geometrical properties of the beam are described by its mode waist $\omega_0$ and its wavelength $\lambda$. The intensity distribution radial to the optical axis is characterized by a Gaussian function with the spot size 

\begin{equation}
\omega(z) = \omega_0 \cdot \sqrt{1+ \left( \frac{z}{z_R} \right)}.
\end{equation}

The wavefront has a spherical shape with a Radius of Curvature (ROC) of

\begin{equation}
R(z)= z + \frac{z_R^2}{z}, 
\end{equation}

where $z_R = \frac{\pi \omega_0^2}{\lambda}$ describes the Rayleigh range \cite{Siegman}. To mode match an optical cavity to a Gaussian beam, the mirror shape needs to match the wavefront at the specific position. Therefore, for  a specific cavity length the mirror structure must have a spherical shape with a corresponding ROC. The profile diameter $d$ of the optic needs to be large enough to cover the intensity profile sufficiently to avoid clipping losses.

In a Fabry-Perot resonator configuration two optical elements trap light inside an optical mode volume $V_m$. This modal volume $V_m$ is ought to be minimized in many quantum optical applications \cite{Ziyun:2012, Benedikter:2017, Gallego:2018}. Consequently optics need to be matched with Gaussian modes down to the size of the used wavelength (as depicted in the inset of figure \ref{MatchedGaussian} b)). For Fabry-Perot cavities mode volumes down to a fraction of $\lambda^3$ are reported to date \cite{Dolan:10, Kelkar_2015}. \\
Applications also demand a high quality factor $Q = \frac{\nu}{\delta \nu}$, which is determined by the ratio between the frequency $\nu$ and the spectral width $\delta \nu$ of a single resonance. A widely used parameter for the characterisation of the optical properties of a resonator is the finesse $F$, which is linked to the quality factor by 

\begin{equation}
Q = \frac{2FL}{\lambda}
\end{equation}

with the optical cavity length $L$ \cite{Hunger2010}. The finesse can be determined by the sum of the round-trip losses via

\begin{equation}
\mathcal{F} = \frac{2\pi}{T_1 + T_2 + \mathcal{S} + \mathcal{A}}
\label{FinesseLosses}
\end{equation}

where T$_1$ and T$_2$ are the transmission values of the individual mirror coatings, $\mathcal{S}$ is the sum of scattering losses of all individual substrate surfaces and $\mathcal{A}$ is the sum of all other losses (e.g. absorption of the mirror coatings). Combining all optical losses with $\mathcal{L}= \mathcal{S} + \mathcal{A}$, the relation $T + R + \mathcal{L} = 1$ must hold for every resonator \cite{Hood2001,Hunger2010,Haubler:19}.\\
Imperfect optics deteriorate the optical properties depending on the scale of the deviation of the surface form an ideal Gaussian wavefront. Large-scaled deviations $\delta \gg \lambda$ from an ideal wavefront lead to the introduction of complex variation of round trip losses and the optical mode profile \cite{Hughes:23}, reducing e.g. the coupling efficiency $\epsilon$ to an ideal Gaussian beam emerging from an optical fibre or the interaction rate to an optical emitter integrated in the cavity. Smaller-scaled imperfections $\delta \approx \lambda$ lead to scattering losses $\mathcal{S}$ \cite{Elson83}, lowering the maximal achievable finesse $\mathcal{F}$. To reduce the latter to a tolerable level, surface qualities with a root-mean-square (RMS) roughness $\sigma_\text{RMS}$ on the (sub-)nanometre scale are favourable. While high quality flat substrates can be achieved through polishing, particular care needs to be taken when fabricating spherical mirrors on the micrometer scale.

\section{Fabrication Process}

\begin{figure}[h]
    \centering
    \def\svgwidth{\columnwidth}
    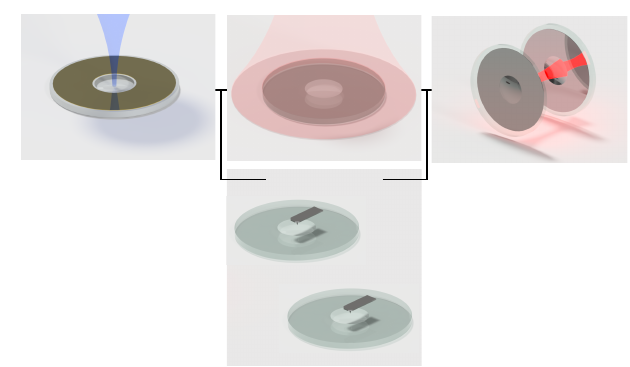
    \caption{Schematic overview of the fabrication (I. - III.) and process control steps (a-d)) of a cavity structure on a fused silica substrate. \textbf{I.} A concave spherical shape is milled into a metal coated fused silica substrate with a focused gallium ion beam (FIB). \textbf{II.} The surface of the structure is smoothed  by application of defocused CO$_2$ laser pulses. \textbf{III.} The glass substrates are coated with a highly reflective mirror coating and placed in a resonator configuration, allowing investigation in terms of cavity properties. \\
\textbf{a)} SEM image of a conical glass substrate with several concave spherical depressions after FIB milling. The green circle marks the spherical structure presented in more detail in this article. \textbf{b)} After chemical removal of residual metal and other contaminants the surface is analysed with an AFM and an interferometric microscope. \textbf{c)} The structure is evaluated again with an AFM and interferometric microscope. \textbf{d)} Light microscope image of the same substrate as in a) after treatment with CO$_2$ laser pulses, AFM investigation and coating with a highly reflective mirror coating at Laseroptik GmbH. The green circle marks the spherical structure presented in more detail in this article.}
    \label{ProcessOverview}
\end{figure}

We demonstrate a fabrication method covering all the above listed requirements, by combining two well established methods, Focused ion beam (FIB) milling and a CO$_2$ laser polishing technique as illustrated in figure \ref{ProcessOverview}. 

In the first step, commercially available fused silica substrates or cleaved surfaces of optical fibres are coated with $\approx 10\,$nm of noble metal (gold or platinum) to establish electrical conductivity on the surface, which is required for the treatment in a FIB. Several sets of concave structures with different geometrical properties (ROC, structure depth $h$ and profile diameter $d$) are then milled into a variety of substrates and optical fibres with a FEI Helios Nanolab 600 SEM+FIB (as seen in fig. \ref{ProcessOverview}.I.). For detailed information about FIB milling parameters and calculations for the realization of concave spherical structures see the supplementary material chapter 1$\,$A and 1$\,$B. Geometrical properties of hemispherical depressions are illustrated in more detail in the supplementary material chapter 3. The remaining metal layer (outside the concave structures) is removed with an aqua regia or Lugol's solution treatment, followed by a chemical cleaning process to remove any remaining contaminants (as described in supplementary Material 1$\,$C). The surface of the structures is then liquified with the help of defocused CO$_2$ laser pulses (fig. \ref{ProcessOverview}.II.), allowing surface tension to eliminate remaining roughness from the FIB milling process. The parameters of the laser pulses (optical power $P$, spot size on the substrate surface $\omega_s$ and pulse length $\tau$) are chosen in a way, that the large-scale shape of the structure remains intact, while small-scale deviations (surface roughness or contaminants) are removed. The process of post FIB treatment (chemical cleaning and smoothing) is monitored by an SEM, an integrated optical microscope, a home built Twyman-Green-Interferometer and a commercial atomic force microscope (AFM) (fig. \ref{ProcessOverview}a)-d)). As a last step assemblies of substrates are coated with two different highly reflective mirror coatings by Laseroptik GmbH (compare to fig. \ref{ProcessOverview}.III.)), before being tested for their optical performance in a cavity configuration. It is noted that examination steps with the AFM (fig. \ref{ProcessOverview}c) and e) are performed for a detailed analysis as presented in this article, which is not necessarily mandatory for process control. Once the smoothing parameters are determined, a light microscope or interferometric setup can be sufficient to ensure elimination of surface roughness and shape preservation (see supplementary material chapter 1.D.).

\section{Structure Evaluation}
\label{eval}

\subsection{Structure Shape}
\label{StrShape}

\begin{figure}[H]
\centering
\includegraphics[scale=1]{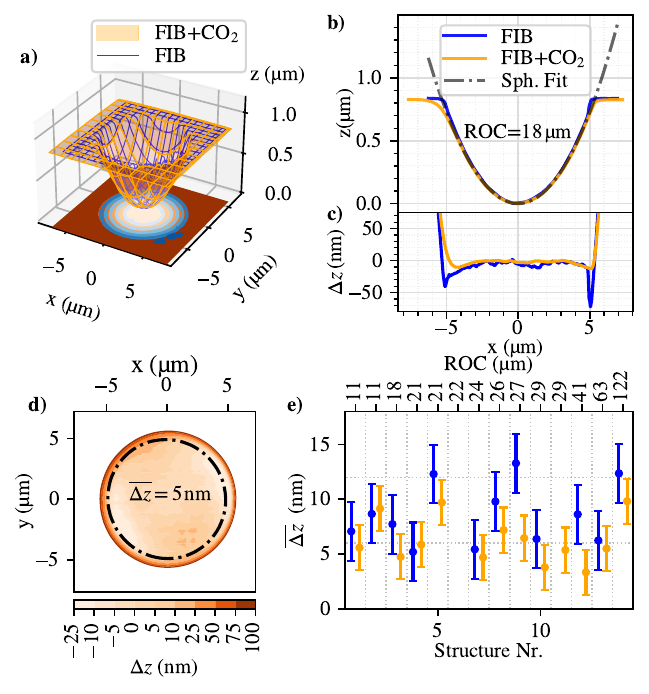}
\caption{\textbf{a)} AFM profiles of an exemplary structure before (blue) and after (orange) surface treatment with defocused CO$_2$ laser pulses ($P=1\,$W, $\omega_s\approx250\,$µm, $\tau_1=41.6\,$ms, $\tau_2=45.5\,$ms). Both profiles are overlapped and a 3D-spherical function is fitted to the data. Differences between the smoothed and un-smoothed structure become only apparent when subtracting them from the spherical fit. \textbf{b)}  Height profile through the center of the structure along the x-axis before (blue) and after smoothing (orange). The profile of the 3D spherical fit is depicted in black. \textbf{c)} Residual deviation from the fitted function before (blue) and after smoothing (orange). The ROC of the structure before and after the smoothing process is 18$\,\mu$m. \textbf{d)} 2D-landscape of the deviation form a spherical shape for the smoothed structure. The average deviation from a sphere of $\Delta z =5\,$nm is calculated for the area inside the dotted line. \textbf{e)} Average deviation from a sphere $\overline{\Delta z}$ for multiple concave structures measured with an AFM before and after application of CO$_2$ laser pulses. In total 14 structures with ROC between 11$\,$µm and 122$\,$µm were investigated, from which all show no increase in deviation from a sphere compared to untreated FIB milled structures. }
\label{3Dshape}
\end{figure}

We overlap both AFM profiles (before and after smoothing) and fit spherical functions respectively. An exemplary set of surface topographies can be seen in figure \ref{3Dshape}$\,$a). We extract the ROC, the depth of the structure $h$, the profile diameter $d$ and the deviation of the profile from a spherical shape $\Delta z$. 
For better visualisation height profiles through the center of the structure (before and after smoothing with CO$_2$ laser pulses) are depicted in figure \ref{3Dshape}$\,$b), including the fitted spherical function. The deviation of the height profile from the spherical fit is represented in figure \ref{3Dshape}$\,$c). We observe reduction of sharp-edged features, as well as small scale surface features in the central region due to the smoothing process. Figure \ref{3Dshape}$\,$d) depicts the deviation from a sphere with regard to the position on the surface. We find an average deviation from a sphere for this particular structure of $\overline{\Delta z}=5\pm6\,$nm excluding the edges of the structure (denoted by black dashed line in fig. \ref{3Dshape}$\,$d)) by averaging over the modulus of the deviation of each data point to the fitted function. 
The area with diameter $d$ is considered the usable surface in a resonator and is considered a hard limit with regards to clipping losses. The diameter $d$ is conservatively estimated so that the usable surface is maximized while the average deviation $\overline{\Delta z}$ is minimized. In supplementary material section 3$\,$a) we estimate the clipping losses of the structure for different cavity lengths.
We perform AFM measurements on a set of 14 structures with different geometrical parameters before and after application of CO$_2$ laser smoothing.
We find that all structures show an equal or reduced average deviation from a sphere, compared to before the smoothing process (as depicted in figure \ref{3Dshape}$\,$e)). To further distinguish between shape preservation and surface smoothing, we performed power spectral density (PSD) calculations of the surface profiles before and after smoothing (as presented in supplementary section 2.E.).\\
For fast process control we implement a home built Twyman-Green interferometer  (compare supplementary section 6.) to determine the surface shape of individual structures before and after individual smoothing pulses \cite{Malacara:2005}. This allows time efficient investigation of the structure shape between individual fabrication steps. While the surface roughness is not resolvable, the smoothing quality can be estimated from a light microscope image once the parameters for the CO$_2$ laser are determined and verified by an AFM (see supplementary section 1.D.). We thereby omit the necessity of time consuming AFM scans for each structure while still achieving high surface quality and structure integrity. \\
The deviation from a sphere in form of an elliptical distortion plays a significant role for the construction of high finesse resonators \cite{Uphoff:2015}.

\begin{figure}[H]
\centering
\def\svgwidth{.75\textwidth}
\includegraphics[scale=1]{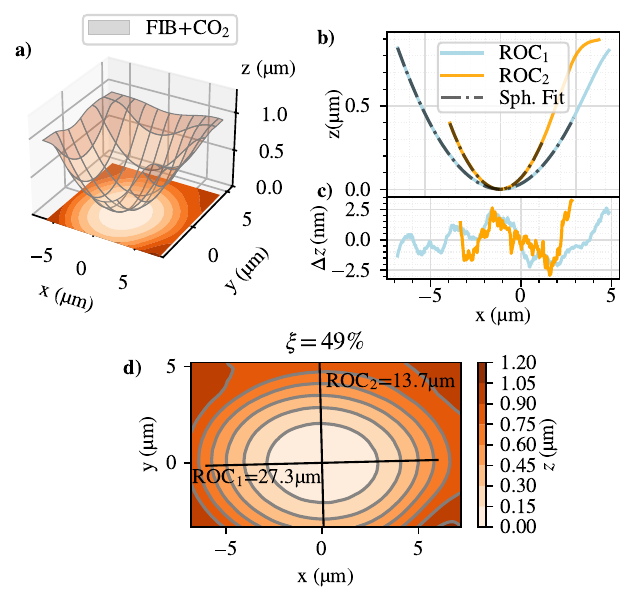}
\caption{\textbf{a)} AFM surface profile of an elliptical structure with ROC$_1$=13.7$\,\mu$m and ROC$_2=$27.3$\,\mu$m after FIB milling and CO$_2$ laser smoothing. \textbf{b)} Height profile along the major (blue) and minor axis (orange) of the ellipsoid with spherical fit (black dashed line). \textbf{c)} Difference of the major and minor axis profile to a circular function. \textbf{d)} Contour profile including the major and minor axis (represented in b)) of the elliptical structure. The ellipticity is found to be $\xi = 49\%$. }
\label{EllipticalStructure}
\end{figure}

An elliptical deformation leads to a frequency splitting between the polarization eigenmodes of the cavity. For our structures the determined deviation from a spherical shape lies already in the range of the resolution limit of the AFM, caused by piezo hysteresis, drifts and levelling errors. Even with these limitations we find values for the ellipticity $\xi$ of only up to a few percent. Interferometric measurements yield values for the ellipticity of below one percent (compare supplementary material section 6.). The cavity measurements presented in chapter \ref{HighFinesse} show no frequency splitting of the polarization eigenmodes which yields an ellipticity $\xi \lessapprox 0.9\% \pm 0.9\%$ (see supplementary material chapter 5.B.). The ellipticity of a cavity structure does not need to be a 
technical limitation but can rather be leveraged as a tool, to  introduce a custom spectral splitting between different polarizations in the resonator (which e.g. matches two distinct optical transitions of one emitter) \cite{Garcia:18}.
We fabricate an exemplary FIB milled and CO$_2$ laser-smoothed structure with a targeted ellipticity of $\xi = 50\%$, where we achieve a final ellipticity of $\xi = 49\%$ (fig. \ref{EllipticalStructure}).

\subsection{Surface Roughness and Scattering Losses}
\label{roughness}

\begin{figure}[H]
\centering
\includegraphics[scale=1]{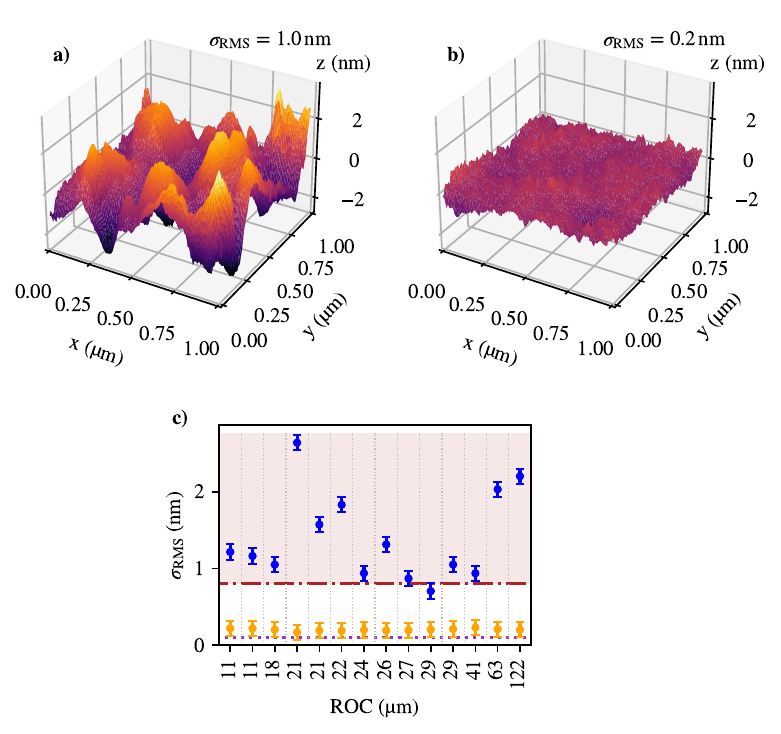}
\caption{\textbf{a)} High resolution AFM scan of the central region of the same FIB milled structure as depicted in blue in figure \ref{3Dshape}. The data is levelled by subtracting a 3rd order polynomial function. The RMS surface roughness is $\sigma_{\text{RMS}} = 1.0\,\pm 0.1\,$nm. \textbf{b)} High resolution AFM scan of the central region of the same structure as in a) after smoothing with CO$_2$ laser pulses (compare to orange surface profile in figure \ref{3Dshape}a) and b)). The RMS surface roughness is reduced to $\sigma_{\text{RMS}} = 0.2\,\pm 0.1\,$nm. \textbf{c)} Measured values for the surface roughness $\sigma_{\text{RMS}}$ of several FIB structures before (blue) and after (orange) treatment with a defocused CO$_2$ laser. While the roughness for FIB structures reaches down to $\sigma_{\text{RMS}}=0.7\pm0.1\,$nm, the CO$_2$ laser treated structures show a consistent roughness of $\sigma_{\text{RMS}}=0.2\,$nm $\pm0.1\,$nm. The red line indicates the limit for the roughness for FIB manufactured structures of $\sigma_\text{RMS}=0.7\,$nm as reported up to date \cite{Trichet15, Walker21}. The noise limit of the AFM for the roughness is calibrated with a Mica Sheet and yields $\sigma_{\text{Noise}}=0.1\,$nm (denoted by a purple dashed line).}
\label{surface}
\end{figure}

We perform high resolution AFM scans in the central region of the structures before and after smoothing to investigate the surface roughness as depicted in fig. \ref{surface} a) and b). We choose a scan range of 1$\,$µm $\times$ 1$\,$µm with a resolution of 2000 $\times$ 2000 pixels. To calibrate our AFM we performed reference measurements on both, Mica Sheets and commercial premium polished optical substrates, yielding $\sigma_{\text{Noise}}=0.1\,$nm (see supplementary material section 2$\,$c)). To achieve maximal comparability of the roughness, the same AFM parameters are used for all scans. The data is levelled by subtracting a 3rd order polynomial function before the root mean square of the roughness is calculated with

\begin{equation}
\sigma_{\text{RMS}} = \sqrt{\frac{1}{N \cdot M} \sum_{n,m=1}^{N,M} (z(x_m,y_n)- \overline{z})^2 }
\label{RoughnessFormula}
\end{equation}

where $\overline{z}$ is the local average height. The roughness from the FIB milling is found to be in the range between $0.7\,$nm and 2.6$\,$nm.

While we observe an increased surface roughness for large structures milled with higher current, we do not find a direct connection between ROC and surface roughness. Larger milled volumes require a higher current to keep a reasonable process time, which increases the minimal spot size of the FIB. At the same time the ion beam is placed on a larger area with the same step interval, making the pattern more prone to small scale defects. The surface roughness is consistently reduced to $\sigma_\text{RMS} = 0.2\pm0.1\,$nm by CO$_2$ laser smoothing independent from structure size and geometry as depicted in fig. \ref{surface} c).\\

\begin{figure}[H]
\centering
\includegraphics[scale=1]{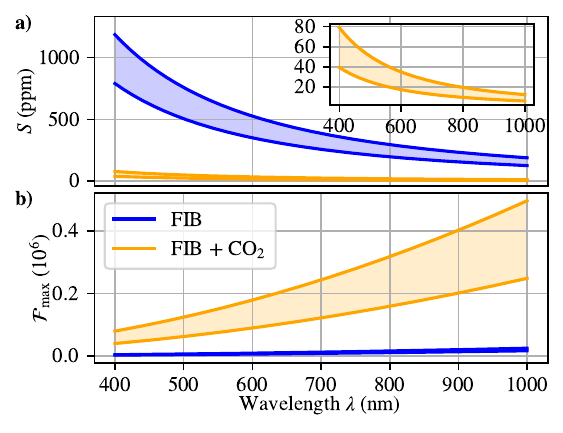}
\caption{\textbf{a)} Estimated scattering losses for the FIB milled structure as depicted in fig. \ref{surface} and \ref{3Dshape}. Losses for untreated FIB structure are represented in blue, for the CO$_2$ laser smoothed structure in orange. The scattering losses are estimated after determining the RMS roughness according to equation \eqref{RoughnessFormula}. \textbf{Inset:} Zoomed representation of the scattering losses for CO$_2$ laser smoothed structures (FIB + CO$_2$). \textbf{b)} Maximal finesse limited by the estimated scattering losses of the surface of the structure for smoothed and untreated FIB structures (neglecting any additional losses, such as mirror coating absorption).}
\label{Scatterlosses}
\end{figure}

For an approximation of scattering losses occurring in the cavity experiment, the scattering coefficient is estimated by

\begin{equation}
\mathcal{S} \approx \left( \frac{4\pi\cdot \sigma_{\text{RMS}}}{\lambda} \right)^2
\label{ScatterEq}
\end{equation}

as used in previous works \cite{Hunger2010}. The estimated losses caused by surface scattering for FIB milled structures with and without treatment of laser pulses are represented in fig. \ref{Scatterlosses}. The maximal achievable finesse is calculated with help of equation \ref{FinesseLosses}. Other optical losses $\mathcal{A}$ like e.g. absorption by the mirror coating or the medium inside the cavity are neglected. As described in section \ref{HighFinesse} the measured losses induced by scattering are lower than predicted by the model presented in fig. \ref{Scatterlosses}, which is why we present as second approach for estimating scattering losses in the supplementary section 4.

\section{Cavity Performance}
\label{Cav}

\subsection{Comparison of FIB and FIB + CO$_2$ Laser}

\begin{figure}[H]
\centering
\includegraphics[scale=1]{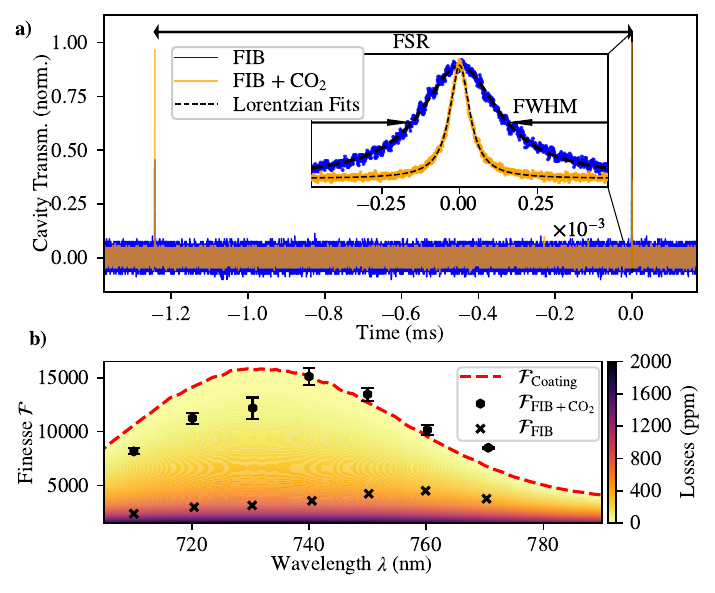}
\caption{\textbf{a)} Laser transmission signal ($\lambda = 740.0\,$nm) of a piezo scanned cavity consisting of a concave and a planar mirror. The transmission signal for the structure presented in section \ref{eval} (ROC = 18$\,$µm) at an average cavity length of $L_\text{Cav} \approx 8\,$µm is represented in orange. A similar structure without treatment of CO$_2$ Laser pulses is depicted in blue. The FSR and and the FWHM of two consecutive cavity resonances are extracted by fitting two Lorentzian functions (black dashed line) to both datasets respectively. A finesse of $\mathcal{F}_\text{FIB + CO2}\approx 16\,000$ is obtained, while the structure without smoothing obtains $\mathcal{F}_\text{FIB} \approx 4100$, corresponding to  losses of $S \approx 600\,$ppm. These losses are  attributed to scattering by a surface with a calculated roughness of $\sigma_{\text{RMS}} = 1.4\,$nm (eq. \eqref{ScatterEq}). \textbf{b)} Measured cavity finesse $\mathcal{F}$ for a FIB structure with application of CO$_2$ laser pulses (same as is in fig. \ref{3Dshape} and \ref{surface}) for different wavelengths $\lambda$. The finesse of a similar FIB structure without application of CO$_2$ laser pulses is depicted as well. The coating limited finesse $\mathcal{F}_\text{Coating}$ ($\mathcal{S}=\mathcal{A}=0$, depicted in red) is calculated by the measured transmission data of the mirror coatings provided by Laseroptik GmbH. The colorbar indicates the reduction in finesse induced by optical losses. The spectral shift between calculated finesse and measured finesse is speculated to arise from deviations in the coating process (as described in supplementary section 9). }
\label{ModFinesse}
\end{figure}

For the comparison between CO$_2$ laser treated and untreated structures we choose a coating with a moderate finesse of $\mathcal{F}\approx 15\,500$ corresponding to a mirror transmission $T=200\,$ppm at a wavelength of 740$\,$nm. We demonstrate that smaller structures, only feasible to fabricate by FIB milling so far, are no longer limited in finesse due to scattering losses. We investigate the optical performance of both structure types by building up a half symmetric test cavity system, where a set of concave structures is placed on a piezo actuator to tune the distance to a macroscopic planar mirror (with equal coating properties). The transmitted and reflected intensity is recorded with an avalanche photo diode (APD) respectively. The finesse 

\begin{equation}
\mathcal{F} = \frac{\text{FSR}}{\text{FWHM}}
\end{equation}

is extracted by the relation between the Free Spectral Range (FSR) and the Full Width at Half Maximum (FWHM) of a resonance. A Lorentzian function is fitted to each resonant peak of the acquired data. We observe a contrast in cavity reflected power of up to $\eta=90\,\%$ (comparing the cavity being on- and off resonant with the laser) for FIB milled and CO$_2$ laser treated structures (as described in the Supplementary Material section 6$\,$b)). From there we estimate the optical coupling efficiency $\epsilon \gtrsim 90\,\%$. A separation of optical mode (mis-)matching from cavity impedance (mis-)matching (ratio between cavity transmission $T$ and losses $\mathcal{L}$), requires more detailed characterization. In particular transmission measurements have to be carried out on the specific experimental apparatus in which the resonator is ultimately used. \cite{Hood2001, Li2006}. \\
With help of equation \eqref{FinesseLosses} the losses caused by scattering are determined. The mirror coating transmissions are specified by Laseroptik GmbH. The cavity transmission signal for the structure presented in the previous sections (ROC = $18\,$µm) is depicted in figure \ref{ModFinesse}$\,$a) (blue) together with the transmission of a similar FIB structure (orange), that has not been treated with a CO$_2$ laser. The respective fitted Lorentzian functions are depicted in black. We acquire cavity transmission data for several wavelengths as represented in fig. \ref{ModFinesse}$\,$b). \\
The FIB structure reaches a maximal finesse of $\mathcal{F}\approx 4100$, corresponding to losses induced by scattering of $\mathcal{S}\approx 600\,$ppm. According to the scattering model of eq. \eqref{ScatterEq}, these losses are caused by a surface roughness of $\sigma_{\text{RMS}}\approx 1.4\,$nm, which is in good accordance to our AFM data. The CO$_2$ laser treated cavity reaches the finesse limited by the coating within the scope of the measurement accuracy ($\Delta \mathcal{F}< 1000$). It is noted that the spectral shift between the measured finesse values and the coating limited finesse $\mathcal{F}_{\text{coating}}$, are speculated to arise from deviations in the mirror coating process (see supplementary material section 9).

\subsection{High Finesse}
\label{HighFinesse}

\begin{figure}[H]
\centering
\includegraphics[scale=1]{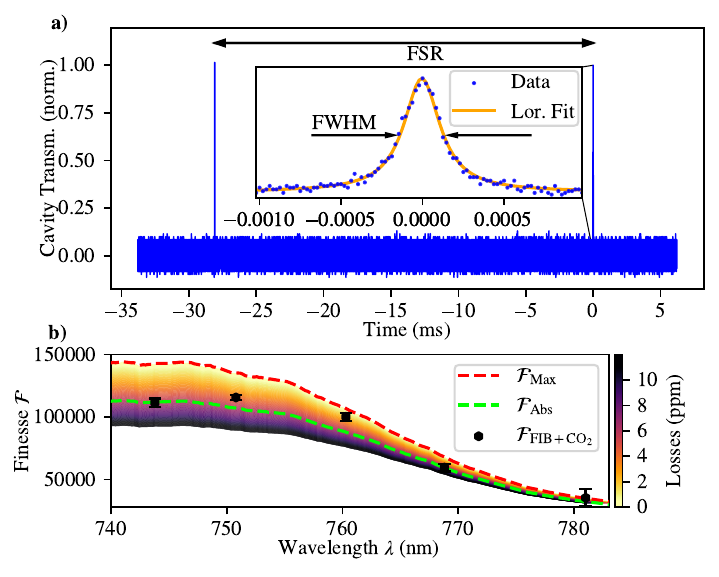}
\caption{\textbf{a)} Laser transmission signal ($\lambda = 750.8\,$nm) of a piezo scanned cavity consisting of two concave mirrors (ROC = 195$\,$µm) at an average cavity length of $L_\text{Cav} = 320\pm10\,\mu$m. The FSR and the FWHM of two consecutive cavity resonances are extracted by fitting two Lorentzian functions (orange line) to both resonances. A finesse of $\mathcal{F}_\text{FIB + CO2}\approx 115\,000$ is obtained. \textbf{b)} Measured finesse for different wavelengths $\lambda$. The finesse limit of the coating neglecting absorptive effects is marked as a red dashed line. The colorbar indicates the reduction in finesse induced by additional optical losses. The green line indicates the maximal finesse considering simulated mirror absorption effects (Laseroptik GmbH), as described in the supplementary material section 8. Effectively reaching the coating limited finesse over a broad spectral range, we conclude that surface scattering effects are negligible compared to absorptive effects of the mirror coating. It is however noted that the simulated absorption values are subject to an uncertainty of a few ppm. Also spectral shifts in the mirror coating have to be taken into account (as described in supplementary material section 8 and 9).}
\label{HighFinesseFigure}
\end{figure}

To analyse the limitations of the here presented technique regarding high finesse applications, we manufacture a set of concave structures with an ROC $\approx 200\,\mu$m (see supplementary material section 5$\,$b)) and have them coated with a highly reflective mirror by Laseroptik GmbH (finesse of a symmetrical cavity $\mathcal{F}_{\text{Max}}\approx 130\,000$, $T\approx 24\,$ppm). We implement a symmetrical cavity consisting out of two (identical) structures at an average cavity length $L_\text{Cav} =320\pm10\,\mu$m. Scanning the cavity length with a piezo element while illuminating the resonator with a laser ($\lambda = 750.8\,$nm) yields a finesse $\mathcal{F}=115\,000\pm3000$ as depicted in figure \ref{HighFinesseFigure}$\,$a). We measure the finesse of the cavity for different wavelengths and compare the values to the coating limit given by the transmission $T(\lambda)$ (specified by Laseroptik GmbH as depicted in supplementary section 8) of the coating while neglecting absorption losses $\mathcal{A}$  as depicted in red in figure \ref{HighFinesseFigure}$\,$b). We include a simulation of the impact of additional losses on the maximally achievable finesse (colorbar). The optical losses introduced by the absorption of the mirror coating material are extrapolated from simulations and measurements performed at Laseroptik GmbH (as described in supplementary section 8). The maximal finesse of the coating, taking absorptive effects into account, is depicted in green. With $\mathcal{A} \approx 6\,$ppm we effectively reach the coating limited finesse over a the designed wavelength range, which indicates that losses from surface scattering are negligible. It is noted that the calculated values for the absorption $\mathcal{A}$ are subject to an uncertainty of few ppm. It is therefore possible that a contribution to the total cavity losses $\mathcal{L}$ from scattering $\mathcal{S}$ still persists. To exclude this contribution fully, more detailed measurements regarding the coating absorption on the concave structures would be required. Spectral shifts of the mirror coating (as described in supplementary section 9) also complicate the determination of the total losses $\mathcal{L}$ in the resonator from the measured finesse values $\mathcal{F}$. \\
We probe the cavity with differently polarized laser light to investigate for a measurable splitting of polarization dependent cavity eigenmodes. The absence of a splitting indicates a high geometrical symmetry of the structure with low ellipticity (as described in supplementary section 3.B.).

\section{Conclusion}

The combination of two mirror fabrication procedures, namely FIB milling and CO$_2$ laser smoothing, allows us to access the advantages of both techniques. The pixel-by-pixel nature of the FIB structuring process enables independent selection of structure parameters with high resolution and over a broad range of parameters. The resolution is up to $\approx$150 times higher as compared to laser dot machining techniques when comparing the resolution of the FIB milling with the spot size of the laser machining setup \cite{Ott:16}. We use that advantage to demonstrate the fabrication of more complex shaped structrures by e.g. introducing a predefined ellipticity. At the same time, the surface quality is at least about 3.5 times better as compared to standard FIB milling when comparing the surface roughness after the FIB milling with the roughness after CO2 laser smoothing \cite{Trichet15}. The treatment with a CO$_2$ laser consistently reduces the surface roughness to $\sigma_{\text{RMS}}=0.2\pm0.1\,$nm independent of the preceding surface quality, which makes the coating the limiting factor for the maximal finesse in most microcavity applications. We demonstrate the fabrication of cavity mirror optics with a finesse of up to $\mathcal{F}= 115\,000$, effectively reaching the coating limit, without generating frequency splitting due to structure deformations or ellipticity. \\
With the repeatable and customizable fabrication of high quality cavity optics we offer a tool-set for a wide range of resonator applications. The option to use the tips of optical fibres as a base substrate, enables the application in spatially restricted experiments.

\section*{Acknowledgements}

The authors gratefully acknowledge support of the Baden-Wuerttemberg Stiftung gGmbH in project AmbientCoherentQE. The authors gratefully acknowledge the funding by the German Federal Ministry of Education and Research (BMBF) within the project QR.X (16KISQ006). J.H.D. and S.R. would like to acknowledge the German Research Foundation (DFG) for financial support within project 416228577 and for support by the Center for Integrated Quantum Science and Technology (IQST). The research of IQST is financially supported by the Ministry of
Science, Research and Arts Baden-Württemberg. Prof. Dr. Christine Kranz, Dr. Gregor Neusser and the Focused Ion Beam Center UUlm are acknowledged for their scientific support during FIB milling. The AFM was funded by the DFG. We thank Prof. Dr. Kay Gottschalk for support. We thank Selene Sachero, Robert Berghaus and Florian Feuchtmayr for technical support. We thank Tobias Gross from Laseroptik GmbH for support regarding coating simulation and coating process. The AFM and interferometric datasets were analysed among others with the open source software Gwyddion \cite{Gwyddion}. 

\section*{Conflict of Interest}
The authors declare no conflict of interest.

\section*{Data availability}
Data underlying the results presented in this paper are not publicly available at this time but may
be obtained from the authors upon reasonable request.

\section*{Supplemental Documents}
See Supplement 1 for supporting content.

\bibliography{bibliography}

\begin{thebibliography}{10}
\newcommand{\enquote}[1]{``#1''}

\bibitem{D2LC00180B}
K.~Malmir, W.~Okell, A.~A.~P. Trichet, and J.~M. Smith,
  \enquote{{Characterization of nanoparticle size distributions using a
  microfluidic device with integrated optical microcavities},}
  {\protect\JournalTitle{Lab Chip}} \textbf{22}, 3499--3507 (2022).

\bibitem{Leroux:10}
I.~D. Leroux, M.~H. Schleier-Smith, and V.~Vuleti\ifmmode~\acute{c}\else
  \'{c}\fi{}, \enquote{{Implementation of Cavity Squeezing of a Collective
  Atomic Spin},} {\protect\JournalTitle{Phys. Rev. Lett.}} \textbf{104}, 073602
  (2010).

\bibitem{Mader2015}
M.~Mader, J.~Reichel, T.~W. H{\"a}nsch, and D.~Hunger, \enquote{A scanning
  cavity microscope,} {\protect\JournalTitle{Nature Communications}}
  \textbf{6}, 7249 (2015).

\bibitem{Hunger:21}
F.~Rochau, I.~S\'anchez~Arribas, A.~Brieussel, \emph{et~al.},
  \enquote{{Dynamical Backaction in an Ultrahigh-Finesse Fiber-Based
  Microcavity},} {\protect\JournalTitle{Phys. Rev. Appl.}} \textbf{16}, 014013
  (2021).

\bibitem{Tenbrake2024}
L.~Tenbrake, A.~Faßbender, S.~Hofferberth, \emph{et~al.}, \enquote{{Direct
  laser-written optomechanical membranes in fiber Fabry-Perot cavities},}
  {\protect\JournalTitle{Nature Communications}} \textbf{15}, 209 (2024).

\bibitem{Haubler:19}
S.~H\"au\ss{}ler, J.~Benedikter, K.~Bray, \emph{et~al.}, \enquote{{Diamond
  photonics platform based on silicon vacancy centers in a single-crystal
  diamond membrane and a fiber cavity},} {\protect\JournalTitle{Phys. Rev. B}}
  \textbf{99}, 165310 (2019).

\bibitem{Jensen_2020}
R.~H\o{}y~Jensen, E.~Janitz, Y.~Fontana, \emph{et~al.},
  \enquote{{Cavity-Enhanced Photon Emission from a Single Germanium-Vacancy
  Center in a Diamond Membrane},} {\protect\JournalTitle{Phys. Rev. Appl.}}
  \textbf{13}, 064016 (2020).

\bibitem{Haubler:21}
S.~Häußler, G.~Bayer, R.~Waltrich, \emph{et~al.}, \enquote{{Tunable
  Fiber-Cavity Enhanced Photon Emission from Defect Centers in hBN},}
  {\protect\JournalTitle{Advanced Optical Materials}} \textbf{9}, 2002218
  (2021).

\bibitem{Bayer2023}
G.~Bayer, R.~Berghaus, S.~Sachero, \emph{et~al.}, \enquote{{Optical driving,
  spin initialization and readout of single SiV centers in a Fabry-Perot
  resonator},} {\protect\JournalTitle{Communications Physics}} \textbf{6}, 300
  (2023).

\bibitem{Feuchtmayr:23}
F.~Feuchtmayr, R.~Berghaus, S.~Sachero, \emph{et~al.}, \enquote{{Enhanced
  spectral density of a single germanium vacancy center in a nanodiamond by
  cavity integration},} {\protect\JournalTitle{Applied Physics Letters}}
  \textbf{123}, 024001 (2023).

\bibitem{Pallmann2024}
M.~Pallmann, K.~Köster, Y.~Zhang, \emph{et~al.}, \enquote{Cavity-{Mediated}
  {Collective} {Emission} from {Few} {Emitters} in a {Diamond} {Membrane},}
  {\protect\JournalTitle{Physical Review X}} \textbf{14}, 041055 (2024).

\bibitem{Casabone2021}
B.~Casabone, C.~Deshmukh, S.~Liu, \emph{et~al.}, \enquote{{Dynamic control of
  Purcell enhanced emission of erbium ions in nanoparticles},}
  {\protect\JournalTitle{Nature Communications}} \textbf{12}, 3570 (2021).

\bibitem{Husel2024}
L.~Husel, J.~Trapp, J.~Scherzer, \emph{et~al.}, \enquote{{Cavity-enhanced
  photon indistinguishability at room temperature and telecom wavelengths},}
  {\protect\JournalTitle{Nature Communications}} \textbf{15}, 3989 (2024).

\bibitem{Zifkin:2024}
R.~Zifkin, C.~D. Rodr\'{\i}guez~Rosenblueth, E.~Janitz, \emph{et~al.},
  \enquote{{Lifetime Reduction of Single Germanium-Vacancy Centers in Diamond
  via a Tunable Open Microcavity},} {\protect\JournalTitle{PRX Quantum}}
  \textbf{5}, 030308 (2024).

\bibitem{Janitz:20}
E.~Janitz, M.~K. Bhaskar, and L.~Childress, \enquote{{Cavity quantum
  electrodynamics with color centers in diamond},}
  {\protect\JournalTitle{Optica}} \textbf{7}, 1232--1252 (2020).

\bibitem{Herrmann_2024}
Y.~Herrmann, J.~Fischer, J.~M. Brevoord, \emph{et~al.}, \enquote{{Coherent
  Coupling of a Diamond Tin-Vacancy Center to a Tunable Open Microcavity},}
  {\protect\JournalTitle{Phys. Rev. X}} \textbf{14}, 041013 (2024).

\bibitem{Englund2007}
D.~Englund, A.~Faraon, I.~Fushman, \emph{et~al.}, \enquote{{Controlling cavity
  reflectivity with a single quantum dot},} {\protect\JournalTitle{Nature}}
  \textbf{450}, 857--861 (2007).

\bibitem{Sanchez:2013}
J.~Miguel-Sánchez, A.~Reinhard, E.~Togan, \emph{et~al.}, \enquote{{Cavity
  quantum electrodynamics with charge-controlled quantum dots coupled to a
  fiber Fabry–Perot cavity},} {\protect\JournalTitle{New Journal of Physics}}
  \textbf{15}, 045002 (2013).

\bibitem{Najer2019}
D.~Najer, I.~S{\"o}llner, P.~Sekatski, \emph{et~al.}, \enquote{{A gated quantum
  dot strongly coupled to an optical microcavity},}
  {\protect\JournalTitle{Nature}} \textbf{575}, 622--627 (2019).

\bibitem{Haas:2014}
F.~Haas, J.~Volz, R.~Gehr, \emph{et~al.}, \enquote{{Entangled States of More
  Than 40 Atoms in an Optical Fiber Cavity},} {\protect\JournalTitle{Science}}
  \textbf{344}, 180--183 (2014).

\bibitem{Rempe:15}
A.~Reiserer and G.~Rempe, \enquote{{Cavity-based quantum networks with single
  atoms and optical photons},} {\protect\JournalTitle{Rev. Mod. Phys.}}
  \textbf{87}, 1379--1418 (2015).

\bibitem{Hartung:2024}
L.~Hartung, M.~Seubert, S.~Welte, \emph{et~al.}, \enquote{A quantum-network
  register assembled with optical tweezers in an optical cavity,}
  {\protect\JournalTitle{Science}} \textbf{385}, 179--183 (2024).

\bibitem{Buifmmode:1997}
V.~Bu\ifmmode~\check{z}\else \v{z}\fi{}ek, G.~Drobn\'y, M.~S. Kim,
  \emph{et~al.}, \enquote{{Cavity QED with cold trapped ions},}
  {\protect\JournalTitle{Phys. Rev. A}} \textbf{56}, 2352--2360 (1997).

\bibitem{Steiner:13}
M.~Steiner, H.~M. Meyer, C.~Deutsch, \emph{et~al.}, \enquote{{Single Ion
  Coupled to an Optical Fiber Cavity},} {\protect\JournalTitle{Phys. Rev.
  Lett.}} \textbf{110}, 043003 (2013).

\bibitem{Ballance:2017}
T.~G. Ballance, H.~M. Meyer, P.~Kobel, \emph{et~al.}, \enquote{{Cavity-induced
  backaction in Purcell-enhanced photon emission of a single ion in an
  ultraviolet fiber cavity},} {\protect\JournalTitle{Phys. Rev. A}}
  \textbf{95}, 033812 (2017).

\bibitem{Birnbaum:2005}
K.~M. Birnbaum, A.~Boca, R.~Miller, \emph{et~al.}, \enquote{{Photon blockade in
  an optical cavity with one trapped atom},} {\protect\JournalTitle{Nature}}
  \textbf{436}, 87--90 (2005).

\bibitem{Puppe:2007}
T.~Puppe, I.~Schuster, A.~Grothe, \emph{et~al.}, \enquote{{Trapping and
  Observing Single Atoms in a Blue-Detuned Intracavity Dipole Trap},}
  {\protect\JournalTitle{Phys. Rev. Lett.}} \textbf{99}, 013002 (2007).

\bibitem{Maunz2004}
P.~Maunz, T.~Puppe, I.~Schuster, \emph{et~al.}, \enquote{Cavity cooling of a
  single atom,} {\protect\JournalTitle{Nature}} \textbf{428}, 50--52 (2004).

\bibitem{Muecke2010}
M.~M{\"u}cke, E.~Figueroa, J.~Bochmann, \emph{et~al.},
  \enquote{{Electromagnetically induced transparency with single atoms in a
  cavity},} {\protect\JournalTitle{Nature}} \textbf{465}, 755--758 (2010).

\bibitem{Kubanek:2011}
A.~Kubanek, M.~Koch, C.~Sames, \emph{et~al.}, \enquote{{Feedback control of a
  single atom in an optical cavity},} {\protect\JournalTitle{Applied Physics
  B}} \textbf{102}, 433--442 (2011).

\bibitem{Kampschulte:2018}
{T. Kampschulte and J. Hecker Denschlag}, \enquote{{Cavity-controlled formation
  of ultracold molecules},} {\protect\JournalTitle{New Journal of Physics}}
  \textbf{20}, 123015 (2018).

\bibitem{Gulati2017}
G.~K. Gulati, H.~Takahashi, N.~Podoliak, \emph{et~al.}, \enquote{Fiber cavities
  with integrated mode matching optics,} {\protect\JournalTitle{Scientific
  Reports}} \textbf{7}, 5556 (2017).

\bibitem{Pfeifer2022}
H.~Pfeifer, L.~Ratschbacher, J.~Gallego, \emph{et~al.}, \enquote{{Achievements
  and perspectives of optical fiber Fabry--Perot cavities},}
  {\protect\JournalTitle{Applied Physics B}} \textbf{128}, 29 (2022).

\bibitem{Brand:2013}
B.~Brandstätter, A.~McClung, K.~Schüppert, \emph{et~al.}, \enquote{Integrated
  fiber-mirror ion trap for strong ion-cavity coupling,}
  {\protect\JournalTitle{Review of Scientific Instruments}} \textbf{84}, 123104
  (2013).

\bibitem{Muller:10}
A.~Muller, E.~B. Flagg, J.~R. Lawall, and G.~S. Solomon,
  \enquote{{Ultrahigh-finesse, low-mode-volume Fabry-Perot microcavity},}
  {\protect\JournalTitle{Opt. Lett.}} \textbf{35}, 2293--2295 (2010).

\bibitem{Hunger:2012}
D.~Hunger, C.~Deutsch, R.~J. Barbour, \emph{et~al.}, \enquote{{Laser
  micro-fabrication of concave, low-roughness features in silica},}
  {\protect\JournalTitle{AIP Advances}} \textbf{2}, 012119 (2012).

\bibitem{Takahashi:14}
H.~Takahashi, J.~Morphew, F.~Oru\v{c}evi\'{c}, \emph{et~al.}, \enquote{{Novel
  laser machining of optical fibers for long cavities with low birefringence},}
  {\protect\JournalTitle{Opt. Express}} \textbf{22}, 31317--31328 (2014).

\bibitem{Ott:16}
K.~Ott, S.~Garcia, R.~Kohlhaas, \emph{et~al.}, \enquote{{Millimeter-long fiber
  Fabry-Perot cavities},} {\protect\JournalTitle{Opt. Express}} \textbf{24},
  9839--9853 (2016).

\bibitem{Hunger2010}
D.~Hunger, T.~Steinmetz, Y.~Colombe, \emph{et~al.}, \enquote{{A fiber
  Fabry–Perot cavity with high finesse},} {\protect\JournalTitle{New Journal
  of Physics}} \textbf{12}, 065038 (2010).

\bibitem{Najer17}
D.~Najer, M.~Renggli, D.~Riedel, \emph{et~al.}, \enquote{{Fabrication of mirror
  templates in silica with micron-sized radii of curvature},}
  {\protect\JournalTitle{Applied Physics Letters}} \textbf{110}, 011101 (2017).

\bibitem{Peng:19}
P.~Qing, J.~Gong, X.~Lin, \emph{et~al.}, \enquote{{A simple approach to
  fiber-based tunable microcavity with high coupling efficiency},}
  {\protect\JournalTitle{Applied Physics Letters}} \textbf{114}, 021106 (2019).

\bibitem{Wachter2019}
G.~Wachter, S.~Kuhn, S.~Minniberger, \emph{et~al.}, \enquote{{Silicon
  microcavity arrays with open access and a finesse of half a million},}
  {\protect\JournalTitle{Light: Science {\&} Applications}} \textbf{8}, 37
  (2019).

\bibitem{Dolan:10}
P.~R. Dolan, G.~M. Hughes, F.~Grazioso, \emph{et~al.}, \enquote{{Femtoliter
  tunable optical cavity arrays},} {\protect\JournalTitle{Opt. Lett.}}
  \textbf{35}, 3556--3558 (2010).

\bibitem{Kelkar_2015}
H.~Kelkar, D.~Wang, D.~Mart\'{\i}n-Cano, \emph{et~al.}, \enquote{{Sensing
  Nanoparticles with a Cantilever-Based Scannable Optical Cavity of Low Finesse
  and Sub-${\ensuremath{\lambda}}^{3}$ Volume},} {\protect\JournalTitle{Phys.
  Rev. Appl.}} \textbf{4}, 054010 (2015).

\bibitem{Trichet15}
A.~A.~P. Trichet, P.~R. Dolan, D.~M. Coles, \emph{et~al.},
  \enquote{{Topographic control of open-access microcavities at the nanometer
  scale},} {\protect\JournalTitle{Opt. Express}} \textbf{23}, 17205--17216
  (2015).

\bibitem{Hughes:23}
W.~J. Hughes, T.~H. Doherty, J.~A. Blackmore, \emph{et~al.}, \enquote{{Mode
  mixing and losses in misaligned microcavities},} {\protect\JournalTitle{Opt.
  Express}} \textbf{31}, 32619--32636 (2023).

\bibitem{Simsek2017}
E.~U. {\c{S}}im{\c{s}}ek, B.~{\c{S}}im{\c{s}}ek, and B.~Orta{\c{c}},
  \enquote{{CO2 laser polishing of conical shaped optical fiber deflectors},}
  {\protect\JournalTitle{Applied Physics B}} \textbf{123}, 176 (2017).

\bibitem{Walker2018}
B.~T. Walker, L.~C. Flatten, H.~J. Hesten, \emph{et~al.},
  \enquote{Driven-dissipative non-equilibrium {Bose}–{Einstein} condensation
  of less than ten photons,} {\protect\JournalTitle{Nature Physics}}
  \textbf{14}, 1173--1177 (2018).

\bibitem{Siegman}
A.~E. Siegman, \emph{{Lasers / Anthony E. Siegman.}} (University Science Books,
  Mill Valley, Calif, 1986).

\bibitem{Ziyun:2012}
Z.~Di, H.~V. Jones, P.~R. Dolan, \emph{et~al.}, \enquote{{Controlling the
  emission from semiconductor quantum dots using ultra-small tunable optical
  microcavities},} {\protect\JournalTitle{New Journal of Physics}} \textbf{14},
  103048 (2012).

\bibitem{Benedikter:2017}
J.~Benedikter, H.~Kaupp, T.~H\"ummer, \emph{et~al.}, \enquote{{Cavity-Enhanced
  Single-Photon Source Based on the Silicon-Vacancy Center in Diamond},}
  {\protect\JournalTitle{Phys. Rev. Appl.}} \textbf{7}, 024031 (2017).

\bibitem{Gallego:2018}
J.~Gallego, W.~Alt, T.~Macha, \emph{et~al.}, \enquote{{Strong Purcell Effect on
  a Neutral Atom Trapped in an Open Fiber Cavity},}
  {\protect\JournalTitle{Phys. Rev. Lett.}} \textbf{121}, 173603 (2018).

\bibitem{Hood2001}
C.~J. Hood, H.~J. Kimble, and J.~Ye, \enquote{Characterization of high-finesse
  mirrors: Loss, phase shifts, and mode structure in an optical cavity,}
  {\protect\JournalTitle{Phys. Rev. A}} \textbf{64}, 033804 (2001).

\bibitem{Elson83}
J.~M. Elson, J.~P. Rahn, and J.~M. Bennett, \enquote{{Relationship of the total
  integrated scattering from multilayer-coated optics to angle of incidence,
  polarization, correlation length, and roughness cross-correlation
  properties},} {\protect\JournalTitle{Appl. Opt.}} \textbf{22}, 3207--3219
  (1983).

\bibitem{Malacara:2005}
{Malacara Z. and Servín M.}, \emph{Interferogram Analysis For Optical Testing
  (2nd ed.).} (2005).

\bibitem{Uphoff:2015}
M.~Uphoff, M.~Brekenfeld, G.~Rempe, and S.~Ritter, \enquote{{Frequency
  splitting of polarization eigenmodes in microscopic Fabry–Perot cavities},}
  {\protect\JournalTitle{New Journal of Physics}} \textbf{17}, 013053 (2015).

\bibitem{Garcia:18}
S.~Garcia, F.~Ferri, K.~Ott, \emph{et~al.}, \enquote{{Dual-wavelength fiber
  Fabry-Perot cavities with engineered birefringence},}
  {\protect\JournalTitle{Opt. Express}} \textbf{26}, 22249--22263 (2018).

\bibitem{Walker21}
B.~T. Walker, B.~J. Ash, A.~A.~P. Trichet, \emph{et~al.}, \enquote{{Bespoke
  mirror fabrication for quantum simulation with light in open-access
  microcavities},} {\protect\JournalTitle{Opt. Express}} \textbf{29},
  10800--10810 (2021).

\bibitem{Li2006}
G.~Li, Y.~Zhang, Y.~Li, \emph{et~al.}, \enquote{Precision measurement of
  ultralow losses of an asymmetric optical microcavity,}
  {\protect\JournalTitle{Appl. Opt.}} \textbf{45}, 7628--7631 (2006).

\bibitem{Gwyddion}
D.~Nečas and P.~Klapetek, \enquote{{Gwyddion: an open-source software for SPM
  data analysis},} {\protect\JournalTitle{Open Physics}} \textbf{10}, 181--188
  (2012).

\end{thebibliography}
\end{document}